\documentclass[pra,aps,twocolumn,amsmath,amssymb,superscriptaddress]{revtex4-1}
\usepackage{epsfig,amsmath}
\usepackage{subfigure}
\usepackage{graphicx}% Include figure files
\usepackage{dcolumn}% Align table columns on decimal point
\usepackage{stmaryrd}
\usepackage{mathrsfs}
\usepackage{pifont}
\usepackage{amsthm}
\usepackage{braket}
\usepackage{amssymb}
\usepackage{bm}
\usepackage{latexsym}
\usepackage[colorlinks=true,linkcolor=blue,citecolor=blue]{hyperref}
\usepackage{color}
\usepackage{epstopdf}
\usepackage[ruled]{algorithm2e}

\begin{document}

\title{Qubit-assisted quantum metrology under a time-reversal strategy}

\author{Peng Chen}
\affiliation{School of Physics, Zhejiang University, Hangzhou 310027, Zhejiang, China}

\author{Jun Jing}
\email{Email address: jingjun@zju.edu.cn}
\affiliation{School of Physics, Zhejiang University, Hangzhou 310027, Zhejiang, China}

\begin{abstract}
We propose a quantum metrology protocol based on a two-step joint evolution of the probe system and an ancillary qubit and quantum measurement. With a proper initial state of the ancillary qubit and an optimized evolution time, the quantum Fisher information (QFI) about the phase parameter encoded in the probe system is found to be determined by the expectation value of the square of a phase generator, irrespective of the probe initial state. Consequently, even if the probe is prepared as a finite-temperature state, faraway from the so-called resource state, e.g., the squeezed spin state or the Greenberger-Horne-Zeilinger state in atomic systems, the QFI in our protocol can approach the Heisenberg scaling $N^2$ with respect to the probe size $N$. This quadratic scaling behavior shows robustness against the imperfections about the initial state of the ancillary qubit and the optimized evolution time of the whole system. Also it is not sensitive to the deviation in system parameters and the qubit decoherence. Using the time-reversal strategy and a single-shot projective measurement, the classical Fisher information (CFI) in our metrology protocol is saturated with its quantum counterpart. Our work thus provides an economical method to reach the Heisenberg limit in metrology precision with no input of entanglement or squeezing.
\end{abstract}

\maketitle

\section{Introduction}

Quantum metrology aims to use various quantum resources including entanglement and squeezing to go beyond the standard quantum limit (SQL), where the uncertainty or fluctuation decreases with $1/\sqrt{N}$, on performing measurements of observable parameters in a probe with $N$ elementary sub-units~\cite{Sun2010fisher,ma2011quantum,genoni2012optical,escher2012quantum,zhong2013fisher,giovannetti2004quantum}. Examples of these parameters are the phases of evolving quantum systems accumulated in multiple scenarios~\cite{Paris2009quantum,helstrom1969quantum,Yurke19862}, such as gravitational wave detection~\cite{Caves1981quantum}, biological sensing~\cite{Taylor2016quantum,Mauranyapin2017evanescent}, atomic clock~\cite{Ludlow2015optical,Katori2011optical}, and magnetometry~\cite{Jones2009magnetic}. A conventional protocol for precision metrology reaching the Heisenberg limit (HL)~\cite{giovannetti2011advances,toth2014quantum,nawrocki2015introduction,polino2020photonic,pezze2018quantum}, where the uncertainty decreases with $1/N$, consists of a prepared probe system [in a maximally correlated state~\cite{bollinger1996optimal}, e.g., the Greenberger-Horne-Zeilinger (GHZ) state in atomic systems and the NOON state in photonic systems], a definite unitary transformation to encode the phase parameter $\theta$ into the probe state, a practical measurement, and a suitable data processing to produce an estimator $\tilde\theta(x)$ about the parameter $\theta$ from the outcome $x$. The estimation precision is quantified by the standard deviation $\delta\theta=\langle[\tilde\theta(x)-\theta]^2\rangle^{1/2}$, which is under the constraint of the Cram\'er-Rao inequality~\cite{helstrom1969quantum,holevo1984probabilistic}: $\delta\theta\ge1/\sqrt{F_c}$ with $F_c$ the classical Fisher information (CFI). Maximizing $F_c$ over all possible measurements gives rise to the quantum Fisher information (QFI) $F_Q$ and hence the quantum Cram\'er-Rao bound (QCRB) $\delta\theta_{\rm min}=1/\sqrt{F_Q}$~\cite{braunstein1994statistical,braunstein1996generalized,luo2003wigner,
pezze2009entanglement,kacprowicz2010experimental} on the attainable sensitivity to estimate $\theta$.

The metrology precision $1/\delta\theta_{\rm min}$ improves with the sub-unit number $N$ of the probe system. With the probe in separable states, we have QFI $\propto N$ for the standard quantum limit. With the probe in maximally entangled states, the QFI scaling can be promoted from linear to quadratic, attaining the Heisenberg scaling $F_Q\propto N^2$. Hence the metrology precision attains HL~\cite{leibfried2004toward,mitchell2004super,boto2000quantum,gerry2000heisenberg}. In atomic systems, so far an $18$-qubit GHZ state has been generated on a quantum processor~\cite{zhang2013quantum,song2019generation,choi2014optimal}. Alternatively, the squeezed spin state can be used to reduce fluctuation to improve metrology precision~\cite{wineland1992spin,kitagawa1993squeezed,wineland1994squeezed,goda2008quantum,
sewell2012magnetic}, which are typically generated by the one-axis twisting (OAT) and two-axis twisting (TAT) interactions~\cite{liu2011spin,zhang2017cavity}. Collective OAT interaction $H_{\rm OAT}\propto J_z^2$ gives rise to a sub-HL noise-reduction $\propto1/N^{2/3}$ for $N$ particles. This interaction is popular in nonlinear platforms, such as Bose-Einstein condensations (BEC)~\cite{gross2010nonlinear,riedel2010atom}, trapped ions~\cite{bohnet2016quantum,lu2019global}, and superconducting qubits~\cite{song2019generation,xu2020probing}. For example, the metrological gain is about $\delta\theta/\delta\theta_{\rm SQL}\sim0.4$ with $120$ atoms~\cite{gross2010nonlinear}. Under the TAT interaction $H_{\rm TAT}\propto J_x^2+J_y^2$, it is theoretically claimed that the squeezing degree can approach HL~\cite{zhang2017cavity,borregaard2017one}. Yet this interaction remains as a challenge in current platforms~\cite{helmerson2001creating,borregaard2017one,macri2020spin}. Quantum nondemolition measurement~\cite{pezze2018quantum,hosten2016measurement,bao2020spin,duan2024continuous} can also generate spin squeezing of atomic ensemble. The metrological gain is found to be $\delta\theta/\delta\theta_{\rm SQL}\sim0.8$ with $10^5$ laser-cooled atoms~\cite{sewell2012magnetic}. In quantum optics, the archetype of a metrology experiment for phase estimation is the Mach-Zehnder interferometer. When beam splitters are replaced with optical parametric amplifiers, the resulting SU(1,1) interferometer has already proven useful in a HL fluctuation reduction only with coherent states~\cite{Yurke19862,li2014phase,li2016phase,jing2011realization,hudelist2014quantum,du20222}. However, the photon loss eventually makes it even below SQL~\cite{marino2012effect}.

Recently, the coupling of the probe with an ancillary system is found to outperform SQL in the parametric estimation. In a protocol for measuring the light rotation with indefinite time direction~\cite{xia2023heisenberg}, an ancillary system serves as a quantum switch leading to a superposition of opposite rotations of the probe system. The rotation precision can thus be enhanced to HL even by using classical probe states. A precision of $12.9$ nrad on the light rotation measurement was experimentally realized with a $150$-order Laguerre-Gaussian beam. In the error-corrected quantum metrology~\cite{demkowicz2014using}, the entanglement between the probe and an ancillary system can be used to reduce noise. Rapid and noiseless quantum operations jointly on the probe and the ancillary system between the encoding rounds empower Ref.~\cite{zhou2018achieving} to perform error correction to suppress the damaging effects of the noise on the probe. By virtue of the time-reversal strategy~\cite{Yurke19862,pezze2017optimal,agarwal2022quantifying,wang2024quantum}, projective measurement can be optimized to saturate QCRB for multiphase estimation~\cite{pezze2017optimal}, given an encoded probe state. The interaction between a photonic system and an ancillary qubit can be manipulated by an external field to implement the unitary transformations along the forward- and reverse-time directions~\cite{luo2023time}. The first generates a photonic entangled NOON state to improve QFI and the second allows CFI to saturate its quantum counterpart.

In this work, we introduce a qubit-assisted metrology protocol to estimate the phase parameter $\theta$ imprinted to the probe system (a spin ensemble) during a spin rotation. It is based on a properly prepared ancillary qubit, the time-reversal strategy, and the projective measurement, but not on any resource state, e.g., the squeezed spin state or GHZ state~\cite{pezze2018quantum}. The QFI of our protocol can be modified from the variance of the phase generator to the mean square of that with respect to the probe state. It therefore shows a Heisenberg scaling in terms of the total spin number even when the probe is prepared as a thermal state. The scaling behavior is not sensitive to the imprecise control over the initial state of the ancillary qubit, the optimized evolution time, and the system parameters. Qubit decoherence reduces QFI yet does not destroy the quadratic scaling. Moreover, CFI of our protocol is found to be coincident with QFI by the time-reversal strategy.

The rest of this work is structured as follows. In Sec.~\ref{sysMod}, we describe the circuit model of our metrology protocol with the ancillary qubit. In Sec.~\ref{conditionTimeRev}, we investigate the conditions about the time-reversal strategy. Detailed derivations about the relevant unitary evolution operator and the phase generator are provided in Appendix~\ref{appendix evolution operator}. In Sec.~\ref{quantumFisherInformation}, we find that QFI in our protocol can be expressed as the mean square of an optimized phase generator with respect to the probe initial state. Detailed derivation about QFI under the deviations in probe-qubit coupling strength and probe eigenfrequency is provided in Appendix~\ref{appendix deviation in parametric control}. In Sec.~\ref{classicalFisherInformation}, we calculate CFI in our protocol as the extractable information from the probability distribution of the output state upon a single-shot projective measurement. In Sec.~\ref{XZInteraction}, we adapt our framework constructed in Secs.~\ref{sysMod}, \ref{conditionTimeRev}, \ref{quantumFisherInformation}, and \ref{classicalFisherInformation} to an alternative interaction between the probe system and an ancillary qubit. In Sec.~\ref{discussion and conclusion}, we discuss the behavior of QFI under the dephasing noise and then summarize the entire work.

\section{Metrology with ancillary qubit}\label{sysMod}

Consider a metrology model consisting of a giant spin probe (spin ensemble) and an ancillary spin-$1/2$. The full Hamiltonian including the free part $H_0$ and the interaction part $H_I$ reads ($\hbar\equiv1$)
\begin{equation}\label{H}
H=H_0+H_I=\omega_PJ_z+\omega_A\sigma_z+gJ_z\sigma_z,
\end{equation}
where $J_\alpha=\sum_{k=1}^{N}\sigma_\alpha^{k}/2$, $\alpha=x,y,z$, denotes the collective spin operator with $\sigma_\alpha^{k}$ the Pauli matrix of the $k$th probe spin, $\omega_P$ and $\omega_A$ represent the energy splitting of the probe spin and the ancillary spin, respectively, and $g$ is the coupling strength between the two components.

The free Hamiltonian and the interaction Hamiltonian in Eq.~(\ref{H}) are feasible in various experiments. In quantum dots~\cite{gillard2022harnessing}, $J_z$ and $\sigma_z$ describe the nuclear spins and the electron spin, respectively. In BEC, the Hamiltonian can be realized by two cavities containing two-component BECs coupled by an optical fiber~\cite{pyrkov2013entanglement}, where the operators $J_z$ and $\sigma_z$ correspond to the Schwinger boson operators $a^\dagger a-b^\dagger b$ in the relevant cavities with $a^\dagger$ and $b^\dagger$ being bosonic creation operators for two orthogonal quantum states. Our interaction Hamiltonian is also similar to that coupling an atomic ensemble to a light beam with the off-resonant Faraday interaction~\cite{bao2020spin,duan2024continuous}, where $J_z$ is the sum of the total angular momentum of individual atoms and $\sigma_z$ denotes the Stokes operator associated with the distinction between the number operators of the photons polarized along orthogonal bases.

\begin{figure}[htbp]
\begin{centering}
\includegraphics[width=0.9\linewidth]{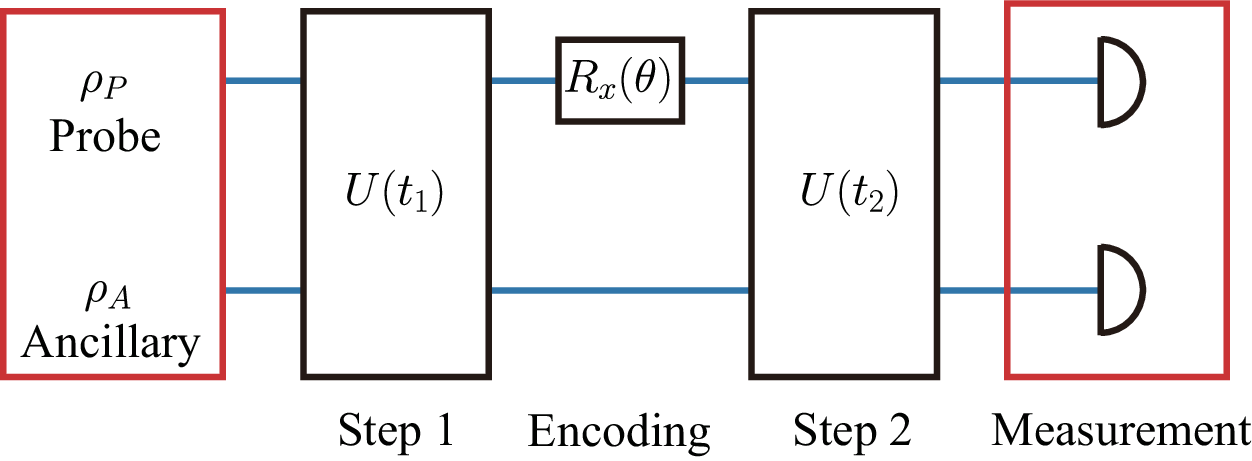}
\caption{Circuit model of our qubit-assisted metrology. Step $1$ and $2$ denote the free joint unitary evolution $U(t_1)$ and its time reversal $U(t_2)$, respectively, in between which a to-be-estimated parameter $\theta$ is encoded into the probe state via a unitary operation $R_x(\theta)$. The output state is determined by a projective measurement on the whole system or the ancillary qubit.}\label{qubit_assisted_protocol}
\end{centering}
\end{figure}

The probe spin ensemble and the ancillary qubit are initially separable, i.e., the input state of the full system is a product state $\rho_P\otimes\rho_A$. It is popular in quantum metrology~\cite{chen2018achieving,chen2018heisenberg} and standard in open-quantum-system dynamics~\cite{Piotr2023}. Practically, it is convenient to obtain an analytical expression for both QFI and CFI with the circuit model in Fig.~\ref{qubit_assisted_protocol}. The entire evolution operator in the circuit can be described by
\begin{equation}\label{U}
U_\theta=U(t_2)R_x(\theta)U(t_1)=e^{-iHt_2}e^{-i\theta J_x}e^{-iHt_1}.
\end{equation}
During Steps $1$ and $2$, the probe system and the ancillary qubit experience a joint time evolution lasting $t_1$ and $t_2$, respectively. Through the parametric encoding of a negligible evolution time (or the full Hamiltonian in between the two steps can be temporarily switched off), we obtain a phase $\theta$ of the probe system by a spin rotation $R_x(\theta)=\exp(-i\theta J_x)$. Experimentally, $\theta$ can be accumulated during the precessing about the $z$ axis~\cite{meyer2001experimental,gross2010nonlinear,ockeloen2013quantum} induced by a certain interaction between the probe and a to-be-measured system. Then the rotation about $x$ axis could be performed by a sequence of $R_x(\theta)=R_y(\pi/2)R_z(\theta)R_y(-\pi/2)$, where $R_y(\pm\pi/2)$ indicates the $\pi/2$ pulse about $y$ axis. The duration time $t_2$ of Step $2$ is determined by $t_1$ of Step $1$ under the time-reversal strategy~\cite{agarwal2022quantifying,wang2024quantum}. On the last stage of the circuit, a projective measurement about the full system is performed on the output state $\rho(\theta)\equiv U_\theta\rho_P\otimes\rho_AU_\theta^\dagger$, whose probability distribution can be used to infer the classical Fisher information about the estimated parameter. When the input state of the probe (spin ensemble) is an eigenstate of $U_\theta$, the entire parameter information can be obtained merely by performing a projective measurement on the ancillary qubit.

It has been shown~\cite{Yurke19862,pezze2017optimal,agarwal2022quantifying,luo2023time,wang2024quantum} that the measurement precision can be certainly enhanced such that CFI approaches QFI when the joint evolution operator $U(t_2)$ becomes the time reversal of $U(t_1)$, i.e.,
\begin{equation}\label{UTimeReversal}
U(t_2)=U^\dagger(t_1).
\end{equation}
It indicates that the full system traces back to the input (separable) state if no phase is encoded.

\section{Conditions on time reversal}\label{conditionTimeRev}

According to the definition of $U_\theta$ in Eq.~(\ref{U}) and the requirement for time reversal in Eq.~(\ref{UTimeReversal}), it is evident that
\begin{equation}\label{UPeriod}
\begin{aligned}
&U_{\theta=0}=e^{-iH(t_1+t_2)}=U(t_1+t_2)=\mathcal I^{2(N+1)} \\
&=\sum_{m=-j}^j|j,m\rangle\langle j,m|\otimes(|e\rangle\langle e|+|g\rangle\langle g|),
\end{aligned}
\end{equation}
where $N=2j$ is the total spin number of the probe ensemble, $\mathcal{I}^n$ is the identity matrix of $n$ dimensions, $|j,m\rangle$ is the eigenstate of the operators $J_z$ with eigenvalue $m$, and $|g\rangle$ and $|e\rangle$ denote the ground and excited states of the ancillary spin, respectively. Using the full Hamiltonian~(\ref{H}), we have
\begin{equation}\label{UHam}
\begin{aligned}
U(t_1+t_2)&=U(T)=\sum_{m=-j}^j|j,m\rangle\langle j,m| \\
&\otimes\left(C_{e,m}|e\rangle\langle e|+C_{g,m}|g\rangle\langle g|\right),
\end{aligned}
\end{equation}
where $C_{e,m}=\exp\{-i[\omega_A+m(\omega_P+g)]T\}$ and $C_{g,m}=\exp\{i[\omega_A-m(\omega_P-g)]T\}$ with $T\equiv t_1+t_2$. When
\begin{equation}\label{CeCg}
C_{e,m}=C_{g,m'},
\end{equation}
Eqs.~(\ref{UPeriod}) and (\ref{UHam}) become equivalent to each other up to a global phase. Consequently, we have
\begin{equation}\label{TCoe}
\left(e^{igT}\right)^{m+m'}\left(e^{i\omega_PT}\right)^{m-m'}=e^{-2i\omega_AT}.
\end{equation}

The parity of the probe-spin number $N$ determines whether the eigenvalue $m$ is an integer or a half-integer. When $N$ is even, $m$ is an integer and then $m\pm m'$ are the same in parity. A sufficient solution for Eq.~(\ref{TCoe}) is thus $e^{igT}=e^{i\omega_PT}=\pm 1$. Straightforwardly, we have
\begin{equation}\label{TConditionEven}
\frac{gT}{\pi}=n_1, \quad \frac{\omega_PT}{\pi}=n_1+2n_2, \quad \frac{\omega_AT}{\pi}=n_3,
\end{equation}
where $n_j$'s, $j=1,2,3$, are proper integers subject to the given magnitudes of $g$, $\omega_P$, and $\omega_A$. When $N$ is odd, $m\pm m'$ are different in parity. The solution can thus be written as
\begin{equation}\label{TConditionOdd}
\frac{gT}{\pi}=n_1, \quad \frac{\omega_PT}{\pi}=n_1+2n_2, \quad \frac{\omega_AT}{\pi}=\frac{n_1}{2}+n_4,
\end{equation}
where $n_j$'s, $j=1,2,4$, are proper integers. For either even or odd $N$, it might be hard to find the exact solution of $T$ if $g$, $\omega_P$, and $\omega_A$ are incommensurable.

The analytical results about $T$ can be confirmed by the numerical simulation over the normalized trace of the unitary operator $U(T)$,
\begin{equation}
\mathcal{F}(T)=\frac{|{\rm Tr}[U(T)]|}{2(N+1)}.
\end{equation}
When $\mathcal{F}(T)=1$, we have $U(T)=\mathcal I^{2(N+1)}$ up to a global phase and vise versa. Once $T$ is specified by a unit trace, the time-reversal condition in Eq.~(\ref{UTimeReversal}) is automatically satisfied by setting $t_2=T-t_1$.

\begin{figure}[htbp]
\begin{centering}
\includegraphics[width=0.9\linewidth]{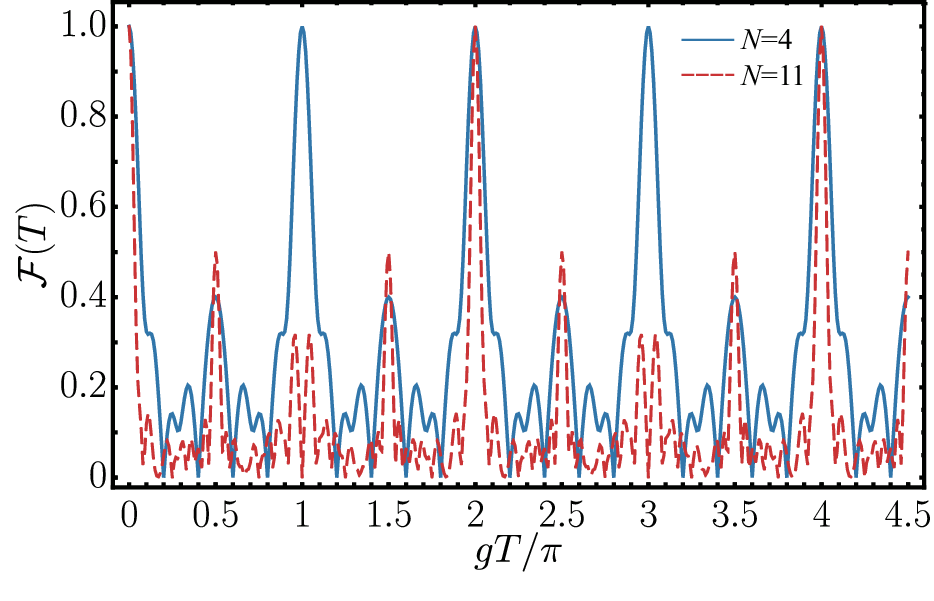}
\caption{Normalized trace $\mathcal{F}(T)$ as a function of the full evolution time $T$ with various spin number $N$. $\omega_P=\omega_A=3g$.}\label{normalized_trace}
\end{centering}
\end{figure}

In Fig.~\ref{normalized_trace}, we demonstrate the normalized trace $\mathcal{F}(T)$ as a function of the evolution time $T$ exemplified with $N=4$ and $N=11$ under a fixed setting that $\omega_P=\omega_A=3g$. The parity of the probe-spin number $N$ determines the time when $\mathcal{F}(T)$ attains unit. For $N=4$ (the even case, see the blue solid line), it is found that the normalized trace attains the maximum value when the full evolution time $gT$ is a multiple of $\pi$. For example, when $gT=\pi$, Eq.~(\ref{TConditionEven}) is valid for $n_1=1$, $n_2=1$, and $n_3=3$. For $N=11$ (the odd case, see the red dashed line), the normalized trace attains unit when $gT$ is a multiple of $2\pi$. Particularly, $gT=2\pi$ is attained by setting $n_1=2$, $n_2=2$, and $n_4=5$ in Eq.~(\ref{TConditionOdd}). Moreover, when both $N$ and $gT/\pi$ are odd integers, one can find that the unitary operator becomes
\begin{equation}\label{UOddHalf}
U(T)=\mathcal I^{N+1}\otimes\sigma_z
\end{equation}
up to a global phase. It will give rise to a vanishing normalized trace $\mathcal{F}(T)=0$. Then the time-reversal strategy is broken. Therefore, choosing a probe ensemble with an even number of spins can significantly accelerate our metrology protocol.

\section{Quantum Fisher information}\label{quantumFisherInformation}

Suppose that we found an exact solution about $T$ on demand of the time-reversal strategy, then the entire unitary evolution operator in Eq.~(\ref{U}) can be rewritten as
\begin{equation}\label{UTimeRevAfter}
U_\theta=e^{iHt_1}e^{-i\theta J_x}e^{-iHt_1}.
\end{equation}
Using the Baker-Campbell-Hausdorff (BCH) formula, we have (the details are provided in Appendix~\ref{appendix evolution operator})
\begin{equation}\label{UTimeRevBCH}
U_\theta=e^{-i\theta\left[\cos(gt_1)J(-\phi)-\sin(gt_1)J\left(\frac{\pi}{2}-\phi\right)\sigma_z\right]},
\end{equation}
where the phase generator is defined as $J(\phi)\equiv\cos(\phi)J_x+\sin(\phi)J_y$ with $\phi=\omega_Pt_1$. We are now on the stage of analyzing the precision limit of the qubit-assisted metrology.

Initially, the ancillary qubit is assumed to be a pure state  $\rho_A=|\varphi\rangle\langle\varphi|$ with~\cite{kitagawa1993squeezed,jin2009spin}
\begin{equation}\label{qubit}
|\varphi\rangle=\cos\left(\frac{\theta_0}{2}\right)|e\rangle+e^{-i\phi_0}\sin\left(\frac{\theta_0}{2}\right)|g\rangle.
\end{equation}
Here $\theta_0$ and $\phi_0$ determine the population imbalance and the relative phase between the two bases, respectively. Without loss of generality, we set the azimuthal angle $\phi_0=0$. For the input state of the probe system, it always can be written as
\begin{equation}\label{probe}
\rho_P=\sum_{i=1}^dp_i|\psi_i\rangle\langle\psi_i|
\end{equation}
in spectral decomposition, where $d$ is the dimension of the density matrix, $p_i$'s are the eigenvalues, and $|\psi_i\rangle$'s are the corresponding eigenstates.

Using Eqs.~(\ref{qubit}) and (\ref{probe}), QFI of the output state for estimating $\theta$ is given by~\cite{braunstein1994statistical,braunstein1996generalized,zhang2013quantum,liu2014quantum}
\begin{equation}\label{QFI}
\begin{aligned}
F_Q &=\sum_{i=1}^d4p_i\langle\psi_i|\langle\varphi|(\partial_\theta U_\theta)^\dagger(\partial_{\theta}U_\theta)|\varphi\rangle|\psi_i\rangle\\
&-\sum_{i,j=1}^d\frac{8p_ip_j}{p_i+p_j}|\langle\psi_i|\langle\varphi|U_\theta^\dagger(\partial_\theta U_\theta)|\varphi\rangle|\psi_j\rangle|^2
\end{aligned}
\end{equation}
with $p_i\neq 0$. It is a difference between two positive terms. The first term is crucial to achieve Heisenberg scaling as shown in the following. A straightforward idea to enhance QFI is thus to minimize the second term, or more generally and precisely, to minimize the expression magnitude inside the absolute value. Using Eq.~(\ref{UTimeRevBCH}), we have
\begin{equation}\label{QFISecondTerm}
\begin{aligned}
\langle\psi_i|\langle\varphi|U_\theta^\dagger(\partial_\theta U_\theta)|\varphi\rangle|\psi_j\rangle=-i\cos(gt_1)\langle\psi_i|J(-\phi)|\psi_j\rangle\\
+i\sin(gt_1)\langle\sigma_z\rangle_\varphi\langle\psi_i|J\left(\frac\pi2-\phi\right)|\psi_j\rangle,
\end{aligned}
\end{equation}
where $\langle\sigma_z\rangle_\varphi=\langle\varphi|\sigma_z|\varphi\rangle$. Under a proper state of the ancillary qubit and an optimized joint-evolution time for Step $1$, i.e.,
\begin{equation}\label{opt}
\theta_0^{\rm opt}=\frac{\pi}{2}, \quad t_{1, {\rm opt}}=\left(n+\frac{1}{2}\right)\frac{\pi}{g},
\end{equation}
with $n$ integer, the two terms in Eq.~(\ref{QFISecondTerm}) can vanish at the same time for arbitrary probe state $|\psi_i\rangle$. Consequently, Eq.~(\ref{UTimeRevBCH}) becomes $U_\theta=\exp(i\theta J_{\rm opt}\sigma_z)$ with $J_{\rm opt}\equiv J(\pi/2-\phi_{\rm opt})$ and $\phi_{\rm opt}\equiv\omega_Pt_{1, {\rm opt}}$.

Equation~(\ref{QFISecondTerm}) indicates the main advantage of our qubit-assisted metrology over a conventional parameter estimation~\cite{giovannetti2011advances,toth2014quantum,nawrocki2015introduction,
polino2020photonic}. The presence of the ancillary qubit provides an extra degree of freedom for control. In the absence of the ancillary qubit, i.e., the coupling strength between the probe system and the ancillary qubit $g=0$, Eq.~(\ref{QFI}) becomes
\begin{equation}\label{QFI_gzero}
\begin{aligned}
F_Q&=\sum_{i=1}^d4p_i\langle\psi _i|J^2(-\phi)|\psi_i\rangle\\
&-\sum_{i,j=1}^d\frac{8p_ip_j}{p_i+p_j}|\langle\psi_i|J(-\phi)|\psi_j\rangle|^2,
\end{aligned}
\end{equation}
where the second term is determined by
\begin{equation}
  \langle\psi_i|J(-\phi)|\psi_j\rangle
=\langle\psi_i|\cos(\omega_Pt_1)J_x-\sin(\omega_Pt_1)J_y|\psi_j\rangle.
\end{equation}
In contrast to Eq.~(\ref{QFISecondTerm}), one can hardly find a state-independent time $t_1$ leading to $\langle\psi_i|J(-\phi)|\psi_j\rangle=0$, simply because $\cos(\omega_Pt_1)$ and $\sin(\omega_Pt_1)$ cannot simultaneously become zero. The only exception is when the probe is initialized as a GHZ state, i.e., $|\psi\rangle=(|j,j\rangle_{\phi}+|j,-j\rangle_{\phi})/\sqrt2$. Here $|j,m\rangle_{\phi}$ with $-j\leq m\leq j$ are the eigenstates of the collective angular momentum operator $J(-\phi)$.

\begin{figure}[htbp]
\begin{centering}
\includegraphics[width=0.9\linewidth]{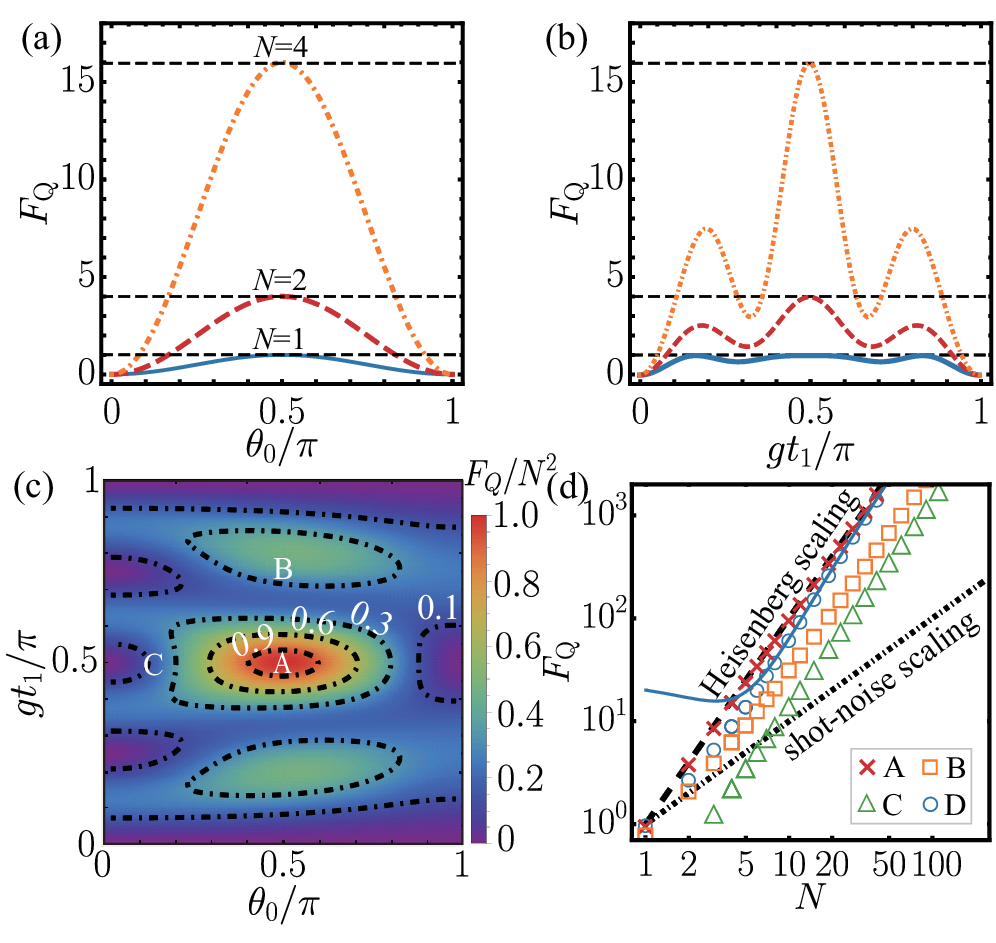}
\caption{(a) QFI as a function of $\theta_0$ with a fixed $gt_1=\pi/2$ and various probe-spin number $N$, (b) QFI as a function of the renormalized evolution time $gt_1$ with a fixed $\theta_0=\pi/2$ and various $N$, (c) Renormalized QFI $F_Q/N^2$ in space of $\theta_0$ and $gt_1$ with $N=4$, and (d) QFI as a function of $N$ for different probe states and parameters. The red crosses indicate point A in (c) for $\rho_P=\rho_P^{\rm opt}=|j,j\rangle_{\rm opt}\langle j,j|$ with $\theta_0=\pi/2$ and $gt_1=\pi/2$. The orange squares indicate point B for $\rho_P^{\rm opt}$ with $\theta_0=\pi/2$ and $gt_1=3\pi/4$. The green triangles denote point C for $\rho_P^{\rm opt}$ with $\theta_0=\pi/8$ and $gt_1=\pi/2$. The blue circles (labeled by D) and the blue solid line are numerical simulation and analytical results for thermal state $\rho_P=\rho_P^{\rm th}$ with $\beta=1$, $\theta_0=\pi/2$, and $gt_1=\pi/2$, respectively. In (a), (b), and (d), the black dashed line indicates the Heisenberg scaling and in (d), the black dot-dashed line indicates the shot-noise scaling. $\omega_P=\omega_A=3g$.}\label{QFI_HL}
\end{centering}
\end{figure}

By Eqs.~(\ref{QFI}) and (\ref{opt}), QFI becomes the mean square of a phase generator with respect to the probe state, i.e.,
\begin{equation}\label{QFISimplify}
F_Q=4\sum_{i=1}^dp_i\langle J_{\rm opt}^2\rangle_{\psi_i},
\end{equation}
which is a weighted average over the expectation value for the operator $J_{\rm opt}^2$ in each eigenstate $|\psi_i\rangle$ of $\rho_P$. The peak value for QFI $F_Q=N^2$ is attainable when the probe system is prepared in a pure state $|j,j\rangle_{\rm opt}$, $|j,-j\rangle_{\rm opt}$, or an arbitrary mixed or superposed state over them, such as $(|j,j\rangle_{\rm opt}\langle j,j|+|j,-j\rangle_{\rm opt}\langle j,-j|)/2$ or $(|j,j\rangle_{\rm opt}+|j,-j\rangle_{\rm opt})/\sqrt2$. Here $|j,m\rangle_{\rm opt}$ with $-j\leq m\leq j$ denote the eigenstates of the optimized collective angular momentum operator $J_{\rm opt}$. Subject to the duration time $t_{1, {\rm opt}}$ and the time-reversal strategy, the bases $|j, m\rangle_{\rm opt}$'s are determined by the magnitudes of $g$, $\omega_P$, and $\omega_A$. Heisenberg scaling can therefore be attainable under optimized $\theta_0$ and $t_1$.

This condition could be confirmed by the numerical simulation in Fig.~\ref{QFI_HL}, where the input state of the spin ensemble is set as $\rho_P^{\rm opt}=|j,j\rangle_{\rm opt}\langle j,j|$ if not otherwise stated. Figure~\ref{QFI_HL}(a) describes the dependence of the quantum Fisher information $F_Q$ on the input parameter of the ancillary qubit $\theta_0$ under an optimized evolution time $t_{1, {\rm opt}}$ in Eq.~(\ref{opt}). It is shown that $F_Q$ follows the Heisenberg scaling when $\theta_0=\pi/2$ for various $N$. In Fig.~\ref{QFI_HL}(b), the initial state of the ancillary qubit is fixed with $\theta_0=\pi/2$ and $F_Q$ arrives at its peak value $F_Q=N^2$ at $gt_1=\pi/2$ as expected by Eq.~(\ref{opt}).

In both Figs.~\ref{QFI_HL}(a) and ~\ref{QFI_HL}(b), $F_Q$ is symmetrical to the optimized points and varies smoothly around them, indicating that our metrology protocol is not sensitive to the deviation in parametric control. If $g\rightarrow g'=g+\Delta g$ and $\omega_P\rightarrow\omega'_P=\omega_P+\Delta\omega_P$ in Eq.~(\ref{QFI}), where $|\Delta gt_{1,{\rm opt}}|\ll1$ and $|\Delta\omega_Pt_{1,{\rm opt}}|\ll1$ indicate the relative deviations in system parameters, then QFI using the optimized probe state, e.g., $\rho_P^{\rm opt}=|j,j\rangle_{\rm opt}\langle j,j|$, becomes
\begin{equation}\label{deviatedFQ}
F_Q\approx N^2-N(N-1)(\Delta\omega_P^2+\Delta g^2)t_{1,{\rm opt}}^2,
\end{equation}
up to the second order of $\Delta g$ and $\Delta\omega_P$. The details are provided in Appendix~\ref{appendix deviation in parametric control}. This result again confirms that our condition in Eq.~(\ref{opt}) is optimal.

In Fig.~\ref{QFI_HL}(c), we take $N=4$ and demonstrate the renormalized QFI $F_Q/N^2$ in the space of $\theta_0$ and $t_1$. The black dot-dashed lines are the contour lines. The central region around point $A$ describes the most optimized condition that is in agreement with Eq.~(\ref{opt}). The surrounding regions about point $B$ and point $C$ describe the sub-optimized conditions with $t_1$ or $\theta_0$ largely departing from Eq.~(\ref{opt}). The behaviors for these points with various $N$ are plotted with the red crosses, the orange squares, and the green triangles, respectively, in Fig.~\ref{QFI_HL}(d). The first one sticks to the upper bound $F_Q=N^2$. Although the rest two are apparently below this bound, an asymptotic behavior to Heisenberg scaling $F_Q\propto N^2$ can appear when $N$ is sufficiently large.

It is interesting to find that QFI could approach the Heisenberg scaling even if the probe ensemble is degraded to be a thermal state in the bases of $|j, m\rangle_{\rm opt}$, i.e.,
\begin{equation}\label{thermal}
\rho_P^{\rm th}=\frac{1}{Z_\beta}e^{-\beta J_{\rm opt}},
\end{equation}
where $Z_\beta={\rm Tr}[\exp(-\beta J_{\rm opt})]$ is the partition function and $\beta$ is the inverse temperature. By Eqs.~(\ref{QFISimplify}) and (\ref{thermal}), we have
\begin{equation}
\begin{aligned}
F_Q&=4\frac{\sum_{m=-N/2}^{N/2}m^2\exp(-m\beta )}{\sum_{m=-N/2}^{N/2}\exp(-m\beta )}\\
&=\frac1{(e^\beta-1)^2}\left[N^2\frac{1-e^{(3+N)\beta }}{1-e^{(1+N)\beta}}\right.\\
&+(N+2)^2\frac{e^{2\beta }-e^{(1+N)\beta}}{1-e^{(1+N)\beta}}\\
&\left.+(N^2+2N-2)\frac{2e^{(2+N)\beta}-2e^\beta}{1-e^{(1+N)\beta}}\right].
\end{aligned}
\end{equation}
Then for a large-size probe, i.e., $N\gg 1$, it is approximated as
\begin{equation}\label{ana thermal}
\begin{aligned}
F_Q&\approx\frac{e^{2\beta}N^2+(N+2)^2-2e^\beta(N^2+2N-2)}{(e^\beta-1)^2}\\
&=N^2-\frac{4}{e^\beta-1}N+4\frac{e^\beta+1}{(e^\beta-1)^2}.
\end{aligned}
\end{equation}
Clearly, $F_Q=N^2$ when $\beta\rightarrow\infty$. It indicates that the Heisenberg-scaling behavior dominates QFI at least in the low-temperature limit. Despite $F_Q$ will become divergent in the high-temperature limit, i.e., $\beta\rightarrow0$, it cannot be used to estimate the metrology precision. The analytical expression in Eq.~(\ref{ana thermal}) is confirmed to match the numerical result in Fig.~\ref{QFI_HL}(d) and both of them (see the blue solid line and the blue circles labeled by $D$) approach the Heisenberg scaling for a sufficient large $N$.

In sharp contrast to the conventional protocol with no ancillary qubit [see, e.g., Eq.~(\ref{QFI_gzero})], our metrology protocol is much broader in optimized probe states. Beyond the GHZ state in an appropriate basis, e.g., $|j, m\rangle_{\phi}$, the Heisenberg limit in metrology precision can be achieved by a product state or even a finite-temperature state.

\section{Classical Fisher information}\label{classicalFisherInformation}

In a practical scenario of parametric estimation, CFI is defined by the amount of information encoded in the probability distribution for the output state~\cite{braun2018quantum,liu2020quantum,tan2021fisher}. In addition, it is upper bounded by its quantum counterpart on the metrology precision. With the time-reversal strategy, it was shown~\cite{Yurke19862,agarwal2022quantifying,wang2024quantum} that CFI can be saturated with QFI, indicating that all the information encoded in the output probability distribution has been extracted. We here derive the CFI of our protocol under the same optimized settings in Eq.~(\ref{opt}) with $n=0$ as for QFI. For the input state $\rho_P\otimes\rho_A$ indicated by Eqs.~(\ref{qubit}) and (\ref{probe}), the output state after the two-step evolution~(\ref{UTimeRevBCH}) reads
\begin{equation}\label{rho}
\begin{aligned}
\rho(\theta)&=U_\theta\rho_P\otimes\rho_AU_\theta^\dagger\\
&=\sum_{m,m'=-j}^j\sum_{i=1}^d\frac{p_ip_{i,m,m'}}{2}|j,m\rangle_{\rm opt}\langle j,m'|\\
&\otimes\left(e^{i\theta m}|e\rangle +e^{-i\theta m}|g\rangle\right)\left(e^{-i\theta m'}\langle e|+e^{i\theta m'}\langle g|\right),
\end{aligned}
\end{equation}
where $p_i$'s are the weights of the eigenstates $|\psi_i\rangle$ in $\rho_P$ and $p_{i,m,m'}\equiv\langle j,m|\psi_i\rangle\langle\psi_i|j,m'\rangle_{\rm opt}$. Subsequently, we perform the projective measurements $|\Psi_{m,\pm}\rangle\langle\Psi_{m,\pm}|$ on the full system as described in Fig.~\ref{qubit_assisted_protocol}, where $|\Psi_{m,\pm}\rangle=|j,m\rangle_{\rm opt}\otimes|\pm\rangle$ with $|\pm\rangle=(|e\rangle\pm|g\rangle )/\sqrt{2}$, the eigenstates of $\sigma_x$ for the ancillary qubit. The probability of detecting the output state in $|\Psi_{m,\pm}\rangle$ reads
\begin{equation}\label{pro}
\begin{aligned}
P(m,\pm|\theta)&=\langle\Psi_{m,\pm}|\rho(\theta)|\Psi_{m,\pm}\rangle\\
&=\frac{1}{4}\sum_{i=1}^dp_ip_{i,m,m}\left\vert e^{i\theta m}\pm e^{-i\theta m}\right\vert^2\\
&=\frac{1\pm\cos(2m\theta)}{2}\sum_{i=1}^dp_ip_{i,m,m},
\end{aligned}
\end{equation}
by which CFI~\cite{braun2018quantum,liu2020quantum,tan2021fisher} can be calculated as
\begin{equation}\label{CFI}
F_c=\sum_{m=-j}^j\left\{\frac{[dP(m,+|\theta)/d\theta]^2}{P(m,+|\theta)}
+\frac{[dP(m,-|\theta)/d\theta]^2}{P(m,-|\theta)}\right\}.
\end{equation}
The first derivative of $P(m,\pm|\theta)$ with respect to the to-be-estimated phase $\theta$ is $dP(m,\pm|\theta)/d\theta=\mp m\sin(2m\theta)\sum_{i=1}^dp_ip_{i,m,m}$ and hence
\begin{equation}\label{dpro}
\frac{[dP(m,\pm|\theta)/d\theta]^2}{P(m,\pm|\theta)}=\frac{2m^2\sin^2(2m\theta)}{1\pm\cos(2m\theta)}\sum_{i=1}^dp_ip_{i,m,m}.
\end{equation}
Substituting Eq.~(\ref{dpro}) to Eq.~(\ref{CFI}), we have
\begin{equation}
\begin{aligned}
F_c=&\sum_{m=-j}^j2m^2\left(\sum_{i=1}^dp_ip_{i,m,m}\right)\sin^2(2m\theta)\\
&\times\left[\frac{1}{1+\cos(2m\theta)}+\frac{1}{1-\cos(2m\theta)}\right]\\
=&4\sum_{m=-j}^j\sum_{i=1}^dm^2p_ip_{i,m,m}\\
=&4\sum_{i=1}^dp_i\langle\psi_i|\left(\sum_{m=-j}^jm^2|j,m\rangle_{\rm opt}\langle j,m|\right)|\psi_i\rangle\\
=&4\sum_{i=1}^dp_i\langle\psi_i|J_{\rm opt}^2|\psi_i\rangle=F_Q.
\end{aligned}
\end{equation}
The last line is exactly the same as Eq.~(\ref{QFISimplify}) under the optimized condition in Eq.~(\ref{opt}), irrespective of the probe state $\rho_P$. It is thus verified that CFI can attain its upper-bound in our metrology. In addition, when the probe is polarized as $\rho_P^{\rm opt}=|j,\pm j\rangle_{\rm opt}\langle j,\pm j|$, Eq.~(\ref{pro}) reduces to
\begin{equation}\label{probality}
P(m,\pm|\theta)=\delta_{m,j}\frac{1\pm\cos(2m\theta)}{2},
\end{equation}
which implies that the magnitude of $\theta$ can be inferred from the probabilities $P(j,+|\theta)$ and $P(j,-|\theta)$. In this case, only performing the projective measurement on the ancillary qubit can yield the Heisenberg-scaling limit $F_c=F_Q=N^2$.

\begin{figure}[htbp]
\begin{centering}
\includegraphics[width=0.9\linewidth]{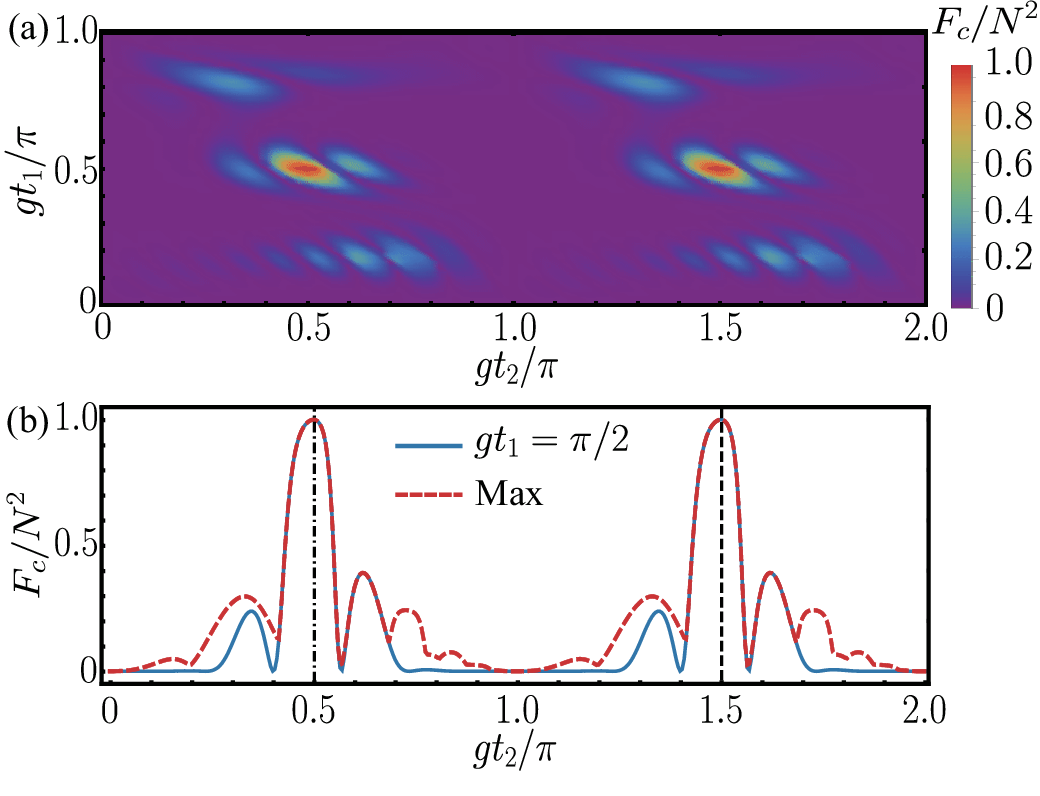}
\caption{(a) Renormalized CFI $F_c/N^2$ in space of $gt_1$ and $gt_2$. (b) Dynamics of the renormalized CFI $F_c/N^2$ versus $gt_2$ with a fixed $gt_1=\pi/2$ (blue solid line) and its maximum value obtained with a varying $gt_1$ (red dashed line). The vertical black dashed line represents the analytical result in Eq.~(\ref{TConditionOdd}). The probe-spin ensemble and the ancillary qubit are initialized in the state $\rho_P^{\rm opt}=|j,j\rangle_{\rm opt}\langle j,j|$ and $|\varphi\rangle=(|e\rangle+|g\rangle)/\sqrt{2}$, respectively. The other parameters are $\phi_{\rm opt}=\omega_P\pi/(2g)$, $N=5$, and $\omega_P=\omega_A=3g$.}\label{CFI_HL}
\end{centering}
\end{figure}

The result about the polarized probe system could be confirmed by the numerical simulation of CFI in Fig.~\ref{CFI_HL}, where the initial state of the ancillary qubit is optimized with $\theta_0=\pi/2$. The dynamics of the output state $\rho(\theta)$ with a general $t_2$ can be obtained by the unitary operator in Eq.~(\ref{U}) and that under the time-reversal strategy is obtained by Eq.~(\ref{rho}). Hence $F_c$ can be numerically evaluated by performing the projective measurements $|\pm\rangle\langle\pm|$ on the ancillary qubit. In Fig.~\ref{CFI_HL}(a), we plot the renormalized CFI $F_c/N^2$ in space of $t_1$ and $t_2$, when the probe spin number is chosen as an odd number $N=5$. It is found that the regions around $gt_1=\pi/2$ [that is in agreement with Eq.~(\ref{opt}) with $n=0$] describe the optimized condition for the Heisenberg-scaling metrology. In addition, it can be more clearly described by the blue solid line in Fig.~\ref{CFI_HL}(b), when we fix $gt_1=\pi/2$. Moreover, the red dashed line indicates the maximum value of CFI with $gt_1$ in the range of $[0, \pi]$. One can find that they periodically match with each other. Note the second maximum point around $gt_2=3\pi/2$ can be predicted by Eq.~(\ref{TConditionOdd}).

CFI can also attain the Heisenberg scaling $F_c=N^2$ when $gt_2=\pi/2$ [see the vertical black dot-dashed line in Fig.~\ref{CFI_HL}(b)]. It is not covered by the sufficient condition for the time-reversal strategy as illustrated by Eq.~(\ref{TConditionOdd}) and Fig.~\ref{normalized_trace}. However, it is straightforward to see that with $gt_1=gt_2=\pi/2$, the entire evolution operator can be written as $U'_{\theta}=U(\pi/g)U^\dagger(t_1){R}_x(\theta)U(t_1)=\sigma_zU_\theta$, where $U_\theta$ is given by Eq.~(\ref{UTimeRevBCH}). Then by using Eq.~(\ref{rho}), we have $\rho'(\theta)=\sigma_zU_\theta\rho_P\otimes\rho_AU_\theta^\dagger\sigma_z=\sigma_z\rho(\theta)\sigma_z$ and hence the probability distribution in Eq.~(\ref{probality}) becomes $P'(j,\pm|\theta)=[1\mp\cos(2m\theta)]/2$, which means $P'(j,\pm|\theta)=P(j,\mp|\theta)$. It indicates that still one can extract all the information encoded in the probe state from the output probabilities and thus $F_c=N^2$.

\section{metrology with $XZ$ interaction}\label{XZInteraction}

The preceding theoretical framework does not confine to the $ZZ$ interaction in the Hamiltonian~(\ref{H}). Our protocol is found to be applicable to another typical spin-spin-bath model~\cite{caldeira1993dissipative,shao1996decoherence,xu2011entanglement,wang2012dynamics} with a Hamiltonian
\begin{equation}\label{Hspinbathmod}
H=H_0+H_I=\omega_PJ_z+\omega_A\sigma_z+gJ_x\sigma_z.
\end{equation}
The relevant circuit model for our metrology in Fig.~\ref{qubit_assisted_protocol} is almost invariant, except that the encoded phase parameter $\theta$ is now obtained by another rotation $R_z(\theta)=\exp(-i\theta J_z)$ of the probe system. Thus the entire unitary evolution is described by
\begin{equation}\label{Uspinbathmod}
U_\theta=U(t_2)R_z(\theta)U(t_1)=e^{-iHt_2}e^{-i\theta J_z}e^{-iHt_1}.
\end{equation}
By using the modified full Hamiltonian~(\ref{Hspinbathmod}), Eq.~(\ref{UHam}) becomes
\begin{equation}\label{UHamspinbathmod}
\begin{aligned}
U(t_1+t_2)&=U(T)=\sum_{m=-j}^j\tilde{C}_{e,m}|j,m\rangle_\phi\langle j,m|\otimes|e\rangle\langle e|\\
&+\sum_{m'=-j}^j\tilde{C}_{g,m'}|j,m'\rangle_{\phi'}\langle j,m'|\otimes|g\rangle\langle g|,
\end{aligned}
\end{equation}
where $\tilde{C}_{e/g,m}=\exp[-i(m\tilde{\omega}\pm\omega_A)T]$ with $\tilde{\omega}\equiv\sqrt{\omega^2_P+g^2}$. Here $|j,m\rangle_\phi$ and $|j,m'\rangle_{\phi'}$ denote the eigenstates of the operators $(\omega_P J_z+g J_x)/\tilde{\omega}$ and $(\omega_P J_z-g J_x)/\tilde{\omega}$ and the relevant eigenvalues are $m$ and $m'$, respectively. According to $U_\theta$ in Eq.~(\ref{Uspinbathmod}) and the time-reversal condition in Eq.~(\ref{UTimeReversal}), we have
\begin{equation}\label{UPeriodspinbathmod}
\begin{aligned}
U_{\theta=0}&=e^{-iH(t_1+t_2)}=U(t_1+t_2)=\mathcal I^{2(N+1)} \\
&=\sum_{m=-j}^j|j,m\rangle_\phi\langle j,m|\otimes|e\rangle\langle e|\\
&+\sum_{m'=-j}^j|j,m'\rangle_{\phi'}\langle j,m'|\otimes|g\rangle\langle g|.
\end{aligned}
\end{equation}
Equation~(\ref{UHamspinbathmod}) becomes equivalent to Eq.~(\ref{UPeriodspinbathmod}) up to a global phase when
\begin{equation}\label{CeCgspinbathmod}
\tilde{C}_{e,m}=\tilde{C}_{g,m'}.
\end{equation}
A sufficient solution for Eq.~(\ref{CeCgspinbathmod}) reads,
\begin{equation}
\frac{\tilde{\omega}T}\pi=2n_5, \quad \frac{\omega_AT}\pi=n_6,
\end{equation}
where $n_j$'s, $j=5,6$, are proper integers. It is independent of the parity of the probe-spin number $N$.

With a proper $T$ on the time-reversal strategy, Eq.~(\ref{Uspinbathmod}) can be rewritten as
\begin{equation}
U_\theta=e^{iHt_1}e^{-i\theta J_z}e^{-iHt_1}.
\end{equation}
By using the BCH formula, we have
\begin{equation}\label{UTimeRevBCHmod}
U_\theta=e^{i\theta(c_zJ_z+c_xJ_x\sigma_z+c_yJ_y\sigma_z)},
\end{equation}
where the coefficients of the phase generators are
\begin{subequations}
\begin{align}\label{cz}
c_z&=-\frac{g^2}{\tilde{\omega}^2}\cos(\tilde{\omega}t_1)-\frac{\omega_P^2}{\tilde{\omega}^2},\\ \label{cx}
c_x&=\frac{g\omega_P}{\tilde{\omega}^2}[\cos(\tilde{\omega}t_1)-1], \\ \label{cy}
c_y&=-\frac{g}{\tilde{\omega}}\sin(\tilde{\omega}t_1).
\end{align}
\end{subequations}
We are now on the stage of minimizing the magnitude of the second term in Eq.~(\ref{QFI}) to ensure the quadratic scaling of QFI. With Eqs.~(\ref{UTimeRevBCHmod}), (\ref{qubit}), and (\ref{probe}), we have
\begin{equation}\label{QFISecondTermMod}
\begin{aligned}
&\langle\psi_i|\langle\varphi|U_\theta^\dagger(\partial_\theta U_\theta)|\varphi\rangle|\psi_j\rangle\\
=&ic_z\langle\psi_i|J_z|\psi_j\rangle+i\langle\sigma_z\rangle_\varphi\langle\psi_i|c_xJ_x+c_yJ_y|\psi_j\rangle.
\end{aligned}
\end{equation}
Under a proper state of the ancillary qubit and an optimized evolution time for Step $1$, i.e.,
\begin{equation}\label{optspinbathmod_strongCoupling}
\theta_0=\frac{\pi}{2}, \quad t_1=\tilde{\omega}^{-1}\left[\arccos\left(-\frac{\omega_P^2}{g^2}\right)+2n\pi\right]
\end{equation}
with an integer $n$, the two terms in Eq.~(\ref{QFISecondTermMod}) can vanish at the same time for arbitrary probe state $|\psi_i\rangle$.

By using Eqs.~(\ref{QFI}) and (\ref{optspinbathmod_strongCoupling}), QFI becomes the mean square of a phase generator, i.e.,
\begin{equation}\label{QFISimplifyMod}
F_Q=4\sum_{i=1}^dp_i\left\langle\left(c_xJ_x+c_yJ_y\right)^2\right\rangle_{\psi_i}.
\end{equation}
It reaches the maximum value $F_Q=N^2$ when the probe spin ensemble is in a pure state $|j, j\rangle'_{\rm opt}$, $|j, -j\rangle'_{\rm opt}$, or an arbitrary mixed or superposed state over them. Here $|j, m\rangle'_{\rm opt}$ with $-j\leq m\leq j$ denotes the eigenstate of the optimized phase generator $c_xJ_x+c_yJ_y$.

\begin{figure}[htbp]
\begin{centering}
\includegraphics[width=0.9\linewidth]{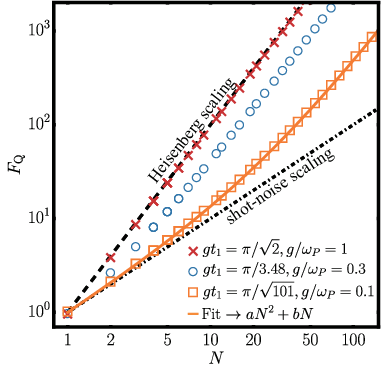}
\caption{QFI as a function of $N$ for $gt_1=\pi/\sqrt{2}$ and $g/\omega_P=1$ (red crosses); for $gt_1=\pi/3.48$ and $g/\omega_P=0.3$ (blue circles); for $gt_1=\pi/\sqrt{101}$ and $g/\omega_P=0.1$ (orange squares). The orange solid line is a fit curve for the last case with $F_Q=aN^2+bN$, where $a\approx0.04$ and $b\approx0.96$. The black dashed line and the black dot-dashed line indicate the Heisenberg scaling and the shot-noise scaling, respectively. }\label{QFI_spinbath}
\end{centering}
\end{figure}

It is evident to see that QFIs in Eqs.~(\ref{QFISimplify}) and (\ref{QFISimplifyMod}) are in the same formation, although with a distinct optimized phase generator determined by the relevant interaction between the probe and ancillary qubit. Then following the same optimized procedure from the output state~(\ref{rho}) to the first derivative of the probability distribution with respect to $\theta$ in Eq.~(\ref{CFI}), one can verify that CFI can attain its upper bound in our metrology, i.e., $F_c=F_Q$, only by replacing $|j, m\rangle_{\rm opt}$ with $|j, m\rangle'_{\rm opt}$. Again, when the probe is prepared as a polarized state $|j,\pm j\rangle'_{\rm opt}\langle j,\pm j|$, the Heisenberg-scaling limit $F_c=F_Q=N^2$ can be conveniently attained by performing the projective measurement $|\pm\rangle\langle\pm|$ on the ancillary qubit.

Since $\arccos(x)$ is defined for $x$ from $-1$ to $1$, the optimal evolution time $t_1$ in Eq.~(\ref{optspinbathmod_strongCoupling}) can only be accessible in the deep strong coupling regime $g\ge\omega_P$. When $g<\omega_P$, only the second term in Eq.~(\ref{QFISecondTermMod}) can vanish for arbitrary probe state and the minimum magnitude of the first term associated with $c_z$ can be found at $\cos(\tilde{\omega}t_1)=-1$ due to Eq.~(\ref{cz}). The sub-optimized condition, that does not render $F_Q=N^2$, can thus be rewritten as
\begin{equation}\label{optspinbathmod_weakCoupling}
\theta_0=\frac{\pi}{2}, \quad t_1=\frac{(2n+1)\pi}{\tilde{\omega}}
\end{equation}
with an integer $n$.

In Fig.~\ref{QFI_spinbath}, we show the dependence of QFI on the probe-spin number $N$ under various $g/\omega_P$, where $t_1$ is specified by Eq.~(\ref{optspinbathmod_strongCoupling}) or Eq.~(\ref{optspinbathmod_weakCoupling}) with $n=0$ and the probe ensemble is initialized as $\rho_P'^{\rm opt}=|j,j\rangle'_{\rm opt}\langle j,j|$. Numerical simulation confirms that when $g=\omega_P$, QFI attains its peak value $F_Q=N^2$ (see the red crosses). The scaling behavior of QFI gradually deviates from the Heisenberg limit with a decreasing $g$. The evolution time $t_1$ and the data for QFI presented by the blue circles and the orange squares are obtained with the sub-optimized condition in Eq.~(\ref{optspinbathmod_weakCoupling}). However, the numerical simulation and fitting for $g/\omega_P=0.1$ (see the orange solid line and the orange squares) indicate that QFI still follow the Heisenberg scaling in an asymptotic way. It is found that $F_Q\approx 0.04N^2+0.96N$, where the two factors are dependent on the ratio $g/\omega_P$ and independent of $N$.

\section{Discussion and Conclusion}\label{discussion and conclusion}

Decoherence can damage any quantum metrology protocol. Here we incorporate the dephasing noise on the ancillary qubit to the ideal model under the Hamiltonian~(\ref{H}). The noise can be described by the following two Kraus operators~\cite{demkowicz2014using}:
\begin{equation}
K_1=\sqrt{1-\frac{x}{2}}\mathcal{I}^2,\quad K_2=\sqrt{\frac{x}{2}}\sigma_z,
\end{equation}
where $x\in[0,1]$ indicates the decoherence rate. The noise-free situation can be recovered when $x=0$. Since the two operators commute with the parametric-encoded unitary evolution operator $U_\theta$ as given in Eq.~(\ref{UTimeRevBCH}), they can be directly inserted into the entire evolution. We then have
\begin{equation}
\begin{aligned}
\rho(\theta)&=U_\theta\rho_P\otimes\left(\sum_{k=1,2}K_k\rho_AK_k^\dagger\right)
U_\theta^\dagger \\ &=U_\theta\rho_P\otimes\rho_A^\prime U_\theta^\dagger.
\end{aligned}
\end{equation}
Under the optimal condition in Eq.~(\ref{opt}), QFI of the output state becomes
\begin{equation}
\begin{aligned}
F_Q&=4\sum_{i=1}^dp_i\langle J_{\rm opt}^2\rangle_{\psi_i}\\
&-\sum_{i,j=1}^d\frac{8x(2-x)p_ip_j}{(2-x)p_i+xp_j}|\langle\psi_i|J_{\rm opt}|\psi_j\rangle|^2.
\end{aligned}
\end{equation}
It is found that the ideal result in Eq.~(\ref{QFISimplify}) can be immediately restored with $x=0$. Under an optimized state $|j,j\rangle_{\rm opt}$ for the probe system in the noise-free situation, we have
\begin{equation}
F_Q=(1-x)^2N^2.
\end{equation}
It turns out that QFI is reduced, yet still follows the Heisenberg scaling behavior $F\propto N^2$ under the dephasing noise.

In summary, we incorporate the joint-evolution-and-quantum-measurement idea to the parameter estimation in our quantum metrology protocol, by which the quantum Fisher information for the phase encoded in the probe system (a spin ensemble) can be optimized to be the mean square of a parametric-dependent phase generator with respect to the probe initial state. It is a metrology protocol for atomic systems with no squeezed or GHZ state and with no nonlinear Hamiltonian. That renders an exact or asymptotic Heisenberg-scaling behavior in the size of the probe system $N$ even when the probe starts from a thermal state. Our calculation shows that the Heisenberg-scaling behavior about metrology precision holds under imperfect controls over the initial state of the ancillary qubit, the optimized evolution time, and the system parameters as long as $N$ is sufficiently large. It even demonstrates certain robustness against the dephasing noise. By virtue of the time-reversal strategy and the projective measurement on the entire system or simply on the ancillary qubit, the classical Fisher information is found to be saturated with its quantum counterpart. Our protocol is applicable to a general spin-spin-bath model, providing insights for enhancing quantum metrology by quantum measurement. In essence, it paves an economical way toward the Heisenberg-scaling metrology.

\section*{Acknowledgments}

We acknowledge financial support from the National Natural Science Foundation of China (Grant No. 11974311).

\appendix
\section{Evolution operator for qubit-assisted circuit under time-reversal strategy}\label{appendix evolution operator}

This appendix derives the unitary evolution operator for the qubit-assisted circuit model in Fig.~\ref{qubit_assisted_protocol}. With the time-reversal condition~(\ref{UTimeReversal}), the entire evolution operator in Eq.~(\ref{UTimeRevAfter}) can be derived as
\begin{equation}
\begin{aligned}
U_\theta&=e^{iHt_1}e^{-i\theta J_x}e^{-iHt_1}\\
&=e^{iHt_1}\left[\sum_{k=0}^\infty\frac1{k!}\left(-i\theta J_x\right)^k\right]e^{-iHt_1}\\
&=e^{iHt_1}e^{-iHt_1}+\left(-i\theta e^{iHt_1}J_xe^{-iHt_1}\right)\\
&+\frac1{2!}\left(-i\theta e^{iHt_1}J_x\right)\left(-i\theta J_xe^{-iHt_1}\right)+\cdots\\
&=\mathcal{I}+\left(-i\theta e^{iHt_1}J_xe^{-iHt_1}\right)\\
&+\frac1{2!}\left(-i\theta e^{iHt_1}J_xe^{-iHt_1}\right)\left(-i\theta e^{iHt_1}J_xe^{-iHt_1}\right)+\cdots\\
&+\frac1{k!}\left(-i\theta e^{iHt_1}J_xe^{-iHt_1}\right)^k+\cdots\\
&=\sum_{k=0}^\infty\frac1{k!}\left(-i\theta e^{iHt_1}J_xe^{-iHt_1}\right)^k\\
&=\exp\left(-i\theta e^{iHt_1}J_xe^{-iHt_1}\right).
\end{aligned}
\end{equation}
in which the operators $e^{\pm iHt_1}$ are put inside the power expansion of the exponential $e^{-i\theta J_x}$.

Using the full Hamiltonian~(\ref{H}) and the commutation relation $[J_z,\sigma_z]=0$,
\begin{equation}\label{jx}
\begin{aligned}
&e^{iHt_1}J_xe^{-iHt_1}\\
=&e^{igt_1J_z\sigma_z}\left(e^{i\omega_Pt_1J_z}J_xe^{-i\omega_Pt_1J_z}\right)e^{-igt_1J_z\sigma_z}\\
=&\cos(\omega_Pt_1)e^{igt_1J_z\sigma_z}J_xe^{-igt_1J_z\sigma_z}\\
-&\sin(\omega_Pt_1)e^{igt_1J_z\sigma_z}J_ye^{-igt_1J_z\sigma_z}.
\end{aligned}
\end{equation}
In the first equivalence, we employ the identity that
\begin{equation}
e^{i\alpha J_z}J_xe^{-i\alpha J_z}=\cos(\alpha)J_x-\sin(\alpha)J_y.
\end{equation}
Subsequently, by using the Baker-Campbell-Hausdorff formula, we have
\begin{equation}\label{BCH}
e^{igt_1J_z\sigma_z}J_xe^{-igt_1J_z\sigma_z}=\sum_{k=0}^{\infty}\frac1{k!}C_{k},
\end{equation}
where $C_{k+1}=igt_1[J_z\sigma_z, C_{k}]$ and $C_0=J_x$. By using the commutation relation $[J_\alpha, J_\beta]=i\epsilon_{\alpha\beta\gamma}J_{\gamma}$ with $\epsilon_{\alpha\beta\gamma}$ the Levi-Civita tensor, we have
\begin{equation}
\begin{aligned}
C_1&=igt_1[J_z\sigma_z, C_0]=-gt_1J_y\sigma_z, \\
C_2&=igt_1[J_z\sigma_z, C_1]=-(gt_1)^2J_x,\\
C_3&=igt_1[J_z\sigma_z, C_2]=(gt_1)^3J_y\sigma_z,
\end{aligned}
\end{equation}
and so on. Equation~(\ref{BCH}) thus becomes
\begin{equation}
\begin{aligned}
&e^{igt_1J_z\sigma_z}J_xe^{-igt_1J_z\sigma_z}\\
=&J_x\sum_{k=0}^{\infty }\frac{(-1)^{k}}{(2k)!}(gt_1)^{2k}-J_y\sigma_z\sum_{k=0}^{\infty }\frac{(-1)^{k}}{(2k+1)!}(gt_1)^{2k+1}\\
=&\cos(gt_1)J_x-\sin(gt_1)J_y\sigma_z.
\end{aligned}
\end{equation}
Similarly, we have
\begin{equation}
e^{igt_1J_z\sigma_z}J_ye^{-igt_1J_z\sigma_z}=\cos(gt_1)J_y+\sin(gt_1)J_x\sigma_z.
\end{equation}
Therefore, Eq.~(\ref{jx}) can be simplified as
\begin{equation}
\begin{aligned}
&e^{iHt_1}J_xe^{-iHt_1}\\
=&\cos(gt_1)[\cos(\omega_Pt_1)J_x-\sin(\omega_Pt_1)J_y]\\
-&\sin(gt_1)[\cos(\omega_Pt_1)J_y+\sin(\omega_Pt_1)J_x]\sigma_z,
\end{aligned}
\end{equation}
and hence
\begin{equation}
U_\theta=e^{-i\theta\left[\cos(gt_1)J(-\phi)-\sin(gt_1)J\left(\frac\pi2-\phi\right)\sigma_z\right]},
\end{equation}
where $J(\phi)\equiv\cos(\phi)J_x+\sin(\phi)J_y$ and $\phi=\omega_Pt_1$. It is exactly Eq.~(\ref{UTimeRevBCH}) in the main text.

\section{QFI under parametric deviation}\label{appendix deviation in parametric control}
This appendix derives quantum Fisher information in the presence of deviations in eigen-frequency of the probe system $\omega_P$ and probe-ancillary qubit coupling strength $g$. By using the proper state of the ancillary qubit and the optimized joint-evolution time given in Eq.~(\ref{opt}), QFI in Eq.~(\ref{QFI}) can be expressed as
\begin{equation}
\begin{aligned}
F_Q&=\sum_{i=1}^d4p_i\langle\cos^2(gt_{1,{\rm opt}})J^2(-\omega_Pt_{1,{\rm opt}})\rangle_{\psi_i}\\
&+\sum_{i=1}^d4p_i\left\langle\sin^2(gt_{1,{\rm opt}})J^2\left(\frac{\pi}{2}-\omega_Pt_{1,{\rm opt}}\right)\right\rangle_{\psi_i}\\
&-\sum_{i,j=1}^d\frac{8p_ip_{j}}{p_i+p_j}|\langle\psi_i|\cos(gt_{1,{\rm opt}})J(-\omega_Pt_{1,{\rm opt}})|\psi_j\rangle|^2.
\end{aligned}
\end{equation}
When $g\rightarrow g'=g+\Delta g$ and $\omega_P\rightarrow\omega_P'=\omega_P+\Delta\omega_P$ with $|\Delta gt_{1,{\rm opt}}|\ll1$ and $|\Delta\omega_Pt_{1,{\rm opt}}|\ll1$ the relative deviations, QFI for the optimized probe state $\rho_P^{\rm opt}=|j,j\rangle_{\rm opt}\langle j,j|$ is
\begin{equation}\label{QFI_expansion}
\begin{aligned}
F_Q&\approx4(1-\Delta g^2t_{1,{\rm opt}}^2)\\
&\times_{\rm opt}\langle j,j|J^2\left[\frac{\pi}{2}-(\omega_P+\Delta\omega_P)t_{1,{\rm opt}}\right]|j,j\rangle_{\rm opt}\\
&+4\Delta g^2t_{1,{\rm opt}}^2\Big\{_{\rm opt}\langle j,j|J^2[-(\omega_P+\Delta\omega_P)t_{1,{\rm opt}}]|j,j\rangle_{\rm opt}\\ &-|_{\rm opt}\langle j,j|J[-(\omega_P+\Delta\omega_P)t_{1,{\rm opt}}]|j,j\rangle_{\rm opt}|^2\Big\}.
\end{aligned}
\end{equation}
Here we use the approximations $\cos^2(g' t_{1,{\rm opt}})\approx\Delta g^2t_{1,{\rm opt}}^2$ and $\sin^2(g' t_{1,{\rm opt}})\approx1-\Delta g^2t_{1,{\rm opt}}^2$. Using Taylor expansion, the three components of QFI in Eq.~(\ref{QFI_expansion}) can be approximated as
\begin{equation}\label{each part QFI}
\begin{aligned}
&_{\rm opt}\langle j,j|J^2\left[\frac{\pi}{2}-(\omega_P+\Delta\omega_P)t_{1,{\rm opt}}\right]|j,j\rangle_{\rm opt}\\
=&\langle j,j|e^{i\Delta\omega_Pt_{1,{\rm opt}}J_y}J_z^2e^{-i\Delta \omega_Pt_{1,{\rm opt}}J_y}|j,j\rangle\\
\approx&j^2-\frac{j(2j-1)}{2}\Delta\omega_P^2t_{1,{\rm opt}}^2,\\
&_{\rm opt}\langle j,j|J^2[-(\omega_P+\Delta\omega_P)t_{1,{\rm opt}}]|j,j\rangle_{\rm opt}\\
=&\langle j,j|e^{i\Delta\omega_Pt_{1,{\rm opt}}J_y}J_x^2e^{-i\Delta\omega_Pt_{1,{\rm opt}}J_y}|j,j\rangle \\
\approx&\frac{j}{2}+\frac{j(2j-1)}{2}\Delta\omega_P^2t_{1,{\rm opt}}^2,\\
&_{\rm opt}\langle j,j|J[-(\omega_P+\Delta\omega_P)t_{1,{\rm opt}}]|j,j\rangle_{\rm opt}\\
=&\langle j,j|e^{i\Delta\omega_Pt_{1,{\rm opt}}J_y}J_xe^{-i\Delta\omega_Pt_{1,{\rm opt}}J_y}|j,j\rangle\\
\approx&j\Delta\omega_Pt_{1,{\rm opt}},
\end{aligned}
\end{equation}
respectively. Substituting Eq.~(\ref{each part QFI}) to Eq.~(\ref{QFI_expansion}), one can obtain QFI up to the quadratic terms about the relative deviations:
\begin{equation}
F_Q\approx N^2-N(N-1)(\Delta\omega_P^2+\Delta g^2)t_1^2.
\end{equation}
Then we obtain Eq.~(\ref{deviatedFQ}).

\bibliographystyle{apsrevlong}
\bibliography{ref}

\end{document}